\documentclass[lettersize,journal]{IEEEtran}
\usepackage{generic}
\usepackage{cite}
\usepackage{amsmath,amssymb,amsfonts}
\usepackage{algorithmic}
\usepackage{graphicx}
\usepackage{algorithm,algorithmic}
\usepackage{hyperref}
\hypersetup{hidelinks=true}
\usepackage{textcomp}
\usepackage{graphicx}
\usepackage{subcaption}

\begin{document}
\title{ARWalker: A Virtual Walking Companion Application}
\author{Pubudu Wijesooriya, Aaron Likens, Nick Stergiou, and Spyridon Mastorakis
\thanks{Date submitted: May 7, 2023. This work is partially supported by the National Science Foundation  (awards CNS-2104700, CNS-2306685, CNS-2016714, and CBET-2124918), ACM SIGMOBILE, and the National Institutes of Health (award NIGMS/P20GM109090). }
\thanks{Pubudu Wijesooriya and Aaron Likens are with the University of Nebraska at Omaha, Omaha, NE 68182, USA (e-mail: pwijesooriya@unomaha.edu, alikens@unomaha.edu). }
\thanks{Nick Stergiou is with the University of Nebraska at Omaha, Omaha, NE 68182, USA and with the Aristotle University of Thessaloniki, Thessaloniki, Greece (email: nstergiou@unomaha.edu, nickstergiou@phed.auth.gr).}
\thanks{Spyridon Mastorakis is with the University of Notre Dame, Notre Dame, IN 46556, USA (email: mastorakis@nd.edu).}}

\maketitle

\begin{abstract}
Extended Reality (XR) technologies, including Augmented Reality (AR), have attracted significant attention over the past few years and have been utilized in several fields, including education, healthcare, and manufacturing. In this paper, we aim to explore the use of AR in the field of biomechanics and human movement through the development of ARWalker, which is an AR application that features virtual walking companions (avatars). Research participants walk in close synchrony with the virtual companions, whose gait exhibits properties found in the gait of young and healthy adults. As a result, research participants can train their gait to the gait of the avatar, thus regaining the healthy properties of their gait and reducing the risk of falls. ARWalker can especially help older adults and individuals with diseases, who exhibit pathological gait thus being more prone to falls. We implement a prototype of ARWalker and evaluate its systems performance while running on a Microsoft Hololens 2 headset. 
\end{abstract}

\begin{IEEEkeywords}
Extended Reality, Gait Rehabilitation, Microsoft Hololens 2, Augmented Reality, Avatars. 
\end{IEEEkeywords}

\section{Introduction}
\label{sec:intro} 

Extended Reality (XR) technologies, including Augmented Reality (AR), Mixed Reality (MR), and Virtual Reality (VR), have attracted significant attention over the past few years~\cite{wijesooriya2023investigating}. XR technologies have been utilized in several fields, including healthcare and rehabilitation, sports, education, training, and manufacturing~\cite{billinghurst2015survey}. At the same time, several XR headsets have been developed, such as the Microsoft Hololens 2 headset, the Meta Oculus Quest and Pro, and the Magic Leap 2. 

In this paper, we explore the use of AR in the field of biomechanics and human movement~\cite{lu2012biomechanics}. Specifically, we developed ARWalker, an AR application that features a virtual walking companion (avatar), so that research participants walk in close synchrony with the gait of the avatar, thus training their gait to the gait of the avatar. According to the optimal movement variability hypothesis~\cite{Optimal-movement-variability}, the gait of young and healthy individuals has been found to exhibit certain properties, while as individuals age or in cases of diseases, it has been found that they exhibit pathological gait~\cite{stergiou2011human}. As a result of that, they are more prone to falls, which can cause severe or even fatal injuries. To this end, the gait of the avatar exhibits properties that have been found in the gait of young and healthy individuals, thus helping research participants regain the healthy properties of their gait and reducing the risk of falls~\cite{Optimal-movement-variability}. 

In other words, ARWalker acts as a gait metronome application, which can be used to carry out gait rehabilitation activities for older and clinical populations. Our contribution in this paper is two-fold: (i) we present the design of ARWalker and the challenges we faced during its design and development process; and (ii) we develop a prototype of ARWalker and we evaluate its systems performance while running on a Hololens 2 headset. 

The rest of this paper is organized as follows. In Section~\ref{sec:background}, we present a brief background on the optimal movement variability hypothesis and prior related work. In Section~\ref{sec:design}, we present the design of ARWalker, and, in Section~\ref{sec:evaluation}, the evaluation of ARWalker. In Section~\ref{sec:discussion}, we discuss various considerations in terms of environmental conditions that we took into account while developing ARWalker and discuss the challenges we faced during its design and development process. Finally, in Section~\ref{sec:conclusion}, we conclude our paper. 
\section{Background and Related Work}
\label{sec:background} 

In this section, we first present a brief background on the optimal movement variability hypothesis and we then discuss prior related work on avatar appearance in VR and gait analysis in XR.

\subsection{Optimal Movement Variability Hypothesis and Gait Metronomes}
When we perform repetitive motor tasks, we produce variations of these tasks.  For example, when we are throwing darts, we cannot hit the same place every time. Similarly, when we are standing still, we constantly sway around a central equilibrium point. Those variations of the motor system are examples of human movement variability, which is considered to be an inherent feature of the human motor system. Furthermore, those variations can be observed in various aspects of movement, such as speed, trajectory, force, and muscle activation patterns. 

The optimal movement variability hypothesis is built upon this knowledge. The hypothesis suggests healthy and natural motor skills are associated with an optimal patterns of movement variability which exhibit characteristics of mathematical chaos \cite{Optimal-movement-variability}. Research suggests pathological movements differ from this optimal patterns in two ways: either an unyielding, unchanging motor patterns while values or motor patterns that are inherently random. 

A gait metronome is a device or method that provides rhythmic stimuli to help individuals maintain a consistent pace or cadence while doing gait exercises. Many types of stimuli can be provided by metronomes, including auditory~\cite{metronomeAuditory}, visual~\cite{metronomeVisual, likens2021irregular}, and tactile~\cite{metronomeVibe}. The emergence of augmented reality provides a great opportunity for metronomes to provide sophisticated and interactive stimuli. In this paper, we explore the direction of realizing a gait metronome that features augmented stimuli implemented in AR. 


\subsection{Relevance of Avatar Appearance in VR}

Numerous studies have been conducted to evaluate the impact of the appearance and attributes of avatars on user behavior and its effects on the physiological and psychological aspects of user immersion. Several studies explored the "Proteus Effect" \cite{proteus:effect} which suggests that the behavior of an individual in a virtual world is changed by the characteristic of their avatar. These studies evaluate how different aspects, such as gender \cite{vr:gender1}, age \cite{vr:age1, vr:age2}, skin color \cite{ vr:skincolor1, vr:skincolor2}, and body types\cite{vr:body1}, induce changes in psychophysical aspects through the illusory ownership of a virtual body. As an example, Jorge Pe{\~n}a et al. examined how the body type of an avatar opponent influences the physical activity of male participants playing an exergame \cite{Pea2016IAW}. Their results showed that participants assigned with normal weight avatar opponents showed more physical activity than the participant assigned with obese avatar opponents.  

Other studies related to avatar appearance investigate the sense of body ownership, avatar visibility, and anthropomorphism. As an example, Ogawa et al. showed that virtual hand realism affects the user's perception of the object sizes \cite{vr:handRealism}. Ogawa et al. also present evidence showing that realistic full-body avatars create a strong sense of embodiment including body ownership \cite{vr:bodyRealism}. Similarly, Choi et al present evidence showing that the naturalness of the glide style improves significantly with avatar head-to-knee visibility \cite{Choiatel}. Interestingly, in an experiment, Oyanagi et al. argued that even nonhuman avatars, such as bird avatars, can change user behavior \cite{oyanagiBird}. They present evidence that becoming a bird avatar can decrease the fear of height in humans. 

In summary, different kinds of research have been performed in the area of avatar representation evaluating many aspects, such as gender, perception, and visibility. This knowledge is useful and has a practical impact on gait rehabilitation. In our work, we incorporate these findings such as full body visibility and natural movement to enhance the sense of embodiment and spatiotemporal feel. 

\subsection{Gait Analysis in XR}

Gait rehabilitation is an important area of health since it is directly correlated with the quality of life. Therefore, lots of research studies have been performed on gait rehabilitation. Almurad et al, for example, observed that when older participants walked in close synchrony with younger participants, this restored the complexity in older participants in three weeks,       and the effect persisted even after two weeks of the training session \cite{arm-in-arm}. The phenomenon behind this result is called complexity matching, which describes interacting systems with similar characteristics aiming to maximize information exchange and attune their complexity to improve coordination \cite{complexity-matching}. 

With the recent advancement of XR, many researchers focused their attention on AR/VR-based gait rehabilitation systems, since they provide a viable and less labor-intensive option for gait rehabilitation. Mohler et al. explored how visual flow influences gait patterns and observed that the visual flow rate has a significant influence on locomotion and preferred walking speed \cite{Mohler2007VisualFI}. Zimmerli et al. took a different approach and developed a VR application that aims to foster motivational aspects in gait rehabilitation \cite{Zimmerli2009VirtualRA}. The application consisted of different environments and tasks. Their results show that such applications lead to higher motivation and increase the activity of patients. Hamzeheinejad and colleagues also took a similar approach and developed a VR application that provides different environments and a virtual trainer for gait exercises \cite{hamzeheinejad2019}. Even though the users assigned less value to the virtual trainer compared to the human trainer, the authors found out that enjoyment was rated higher in the VR setup compared to the non-VR setup. Furthermore, they did not find any negative side effects in their experiments. 

In addition to the above work, there are many other approaches that combined gait rehabilitation and VR \cite{other-Gait1, other-Gait2, other-Gait3, other-Gait4, other-Gait5, other-Gait6}. Most of these approaches use VR with combinations of sensors, treadmills, or some form of specialized hardware such as Locomat.  However, in line with our approach, Held et al. developed the ARISE (Augmented Reality for gait Impairments after StrokE) system, which uses Hololens 2 with a sensor-based motion capture system \cite{Held2020AugmentedRR}. The ARISE system provides a personalized training environment with patient-centered feedback and a monitoring system and has shown promising results in gait improvements in a 74 years man who had a right-hemispheric ischemic stroke in the thalamus, capsula interna, and right temporal lobe 7 years before participating in the experiment.

Even though ARISE and our work have similarities, our system is different than ARISE. For example, ARISE system uses an AR-based parkour course that has an area with visualizations of real-life obstacles and barriers to provide gait rehabilitation. Our system uses a different approach than ARISE. Instead of a parkour course, our system exploits complexity matching \cite{arm-in-arm} and proteus effects \cite{proteus:effect} to provide gait rehabilitation. In our system, users spawn an avatar with healthy complexity in front of them and restores the complexity of their own gait by walking beside the avatar. One of the main advantages of our system compared to other existing systems is its mobility. Our system offers more freedom compared to the other systems such as treadmill-based VR training \cite{other-Gait4} or robot-assisted VR training \cite{other-Gait6}. The user can simply put on the AR headset and start the exercise in both indoor and outdoor environments. Unlike other systems, our system augments the real world with virtual elements (avatars), providing opportunities for users to interact with the real world. Finally, we incorporate many of the findings of previous work, such as the importance of the avatar appearance, avatar age, skin color, and realism by creating a diverse set of avatars that can simulate natural gait patterns.

\section{ARWalker Design}
\label{sec:design} 

In this section, we discuss the design of ARWalker.

\subsection{ARWalker Operation}

\begin{figure*}[!t]

	\centering
	\begin{subfigure}{0.45\textwidth}
		\centering
		\includegraphics[width=1\textwidth, height=6.2cm]{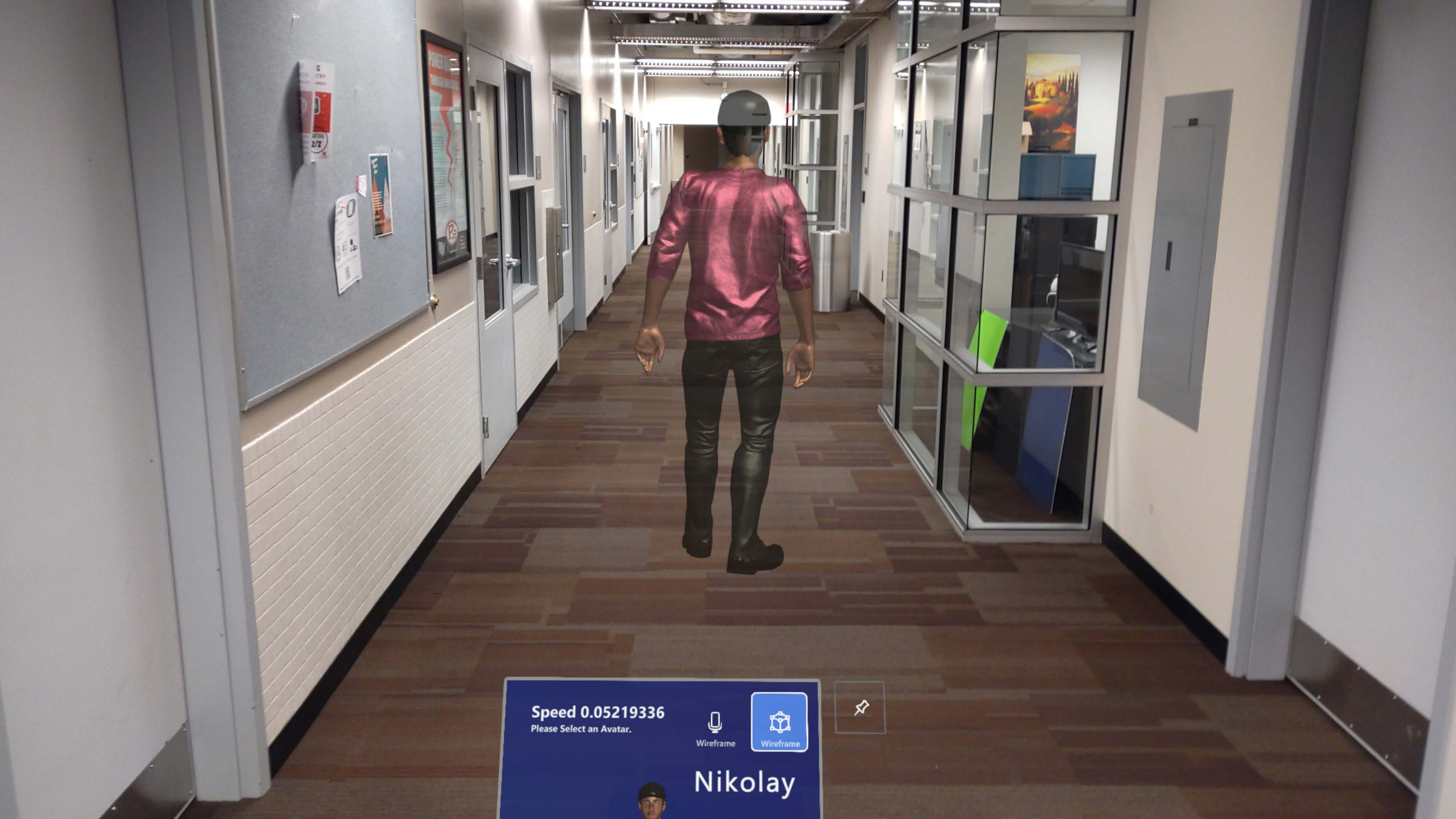}
		\caption{An example of a male avatar walking in front of the user's field of view.} \hfill
		\label{Figure:avatarfemalewalking}
	\end{subfigure} 
	\hfill
	\begin{subfigure}{0.45\textwidth}
		\centering
		\includegraphics[width=1\textwidth]{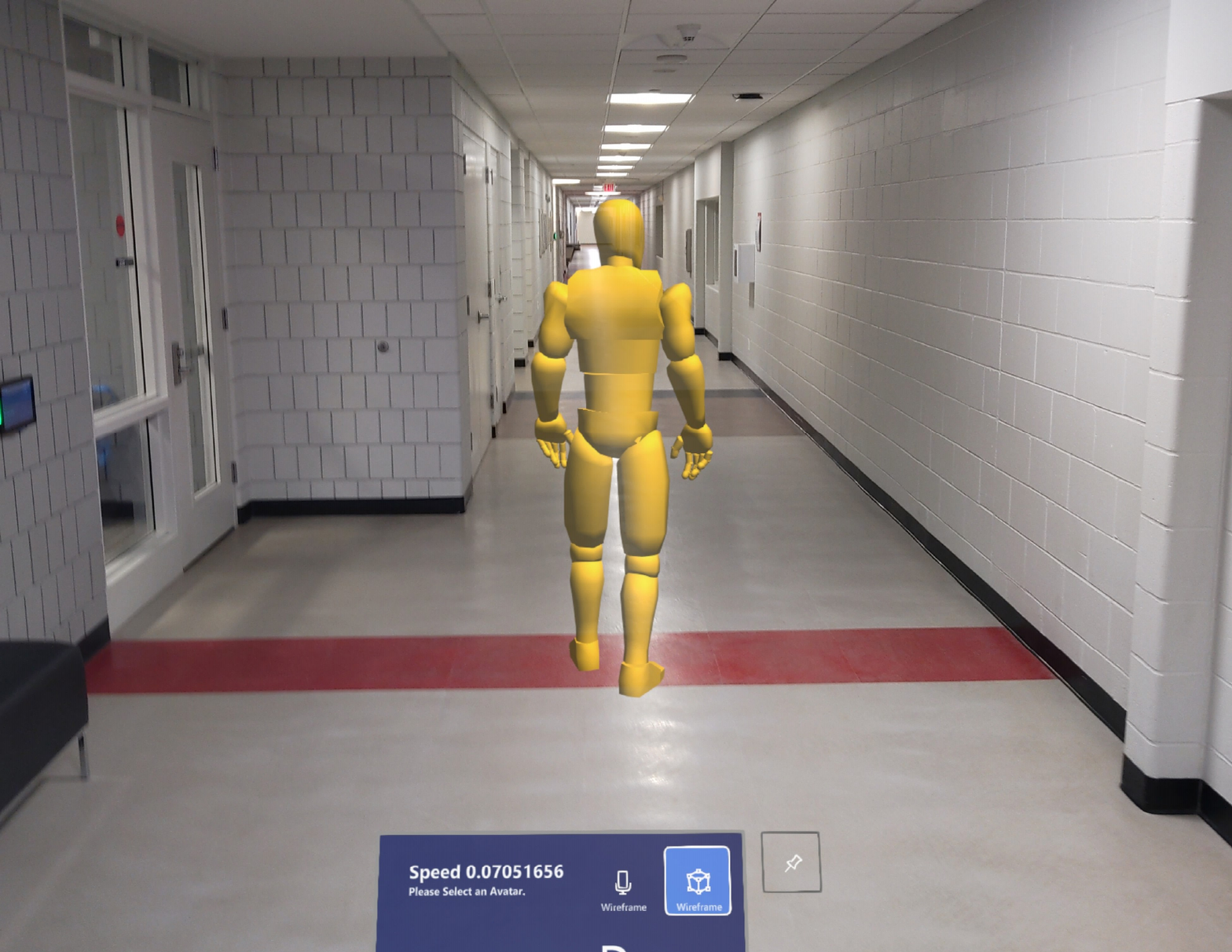}
		\caption{An example of a humanoid avatar walking in front of the user's field of view. }  \hfill
		\label{Figure:avatarmalewalking}
	\end{subfigure}
	\caption{Different avatars walking in front of the user's field of view.}
    \label{figure:avatars}
	\vspace{-0.2cm}
\end{figure*}

Our metronome application run on the Hololens 2 headset (Section~\ref{subsec:headset}). On a high level, it works as follows. At the start of the application, it allows users to choose an avatar from a diverse pool of avatars (Section~\ref{subsec:avatarselection}) and configure various parameters of the metronome operation through UIs we have developed (Section~\ref{subsec:ui}). In Figure~\ref{figure:avatars}, we present the operation of our avatar-based metronome with examples of male and humanoid avatars from our avatar pool. After that, the selected avatar will spawn in front of the user. 
When the user starts walking, the selected avatar will walk in front of the user staying in the user's field of view. Our hypothesis is that through context matching the user will restore the complexity of their own gait by synchronizing their gait to the gait of the avatar (Sections~\ref{subsec:noises} and~\ref{subsec:coupling}).

We use a solver handler function in MRTK to track the head position of a user. By using head movements, the avatar turns in the same way as the user's head. We also implemented a singleton instance that calculates the user's speed, distance, and rotations. Whenever the user starts to move, this signals the animation to start and transition the avatar from idle to a walking animation. This approach allows the user to navigate in any environment in a lightweight manner in terms of the occupied resources (CPU and memory usage) of the headset.

\subsection{Apparatus}
\label{subsec:headset}

To combine the virtual elements with the real world, we used Microsoft Hololens 2 with eye tracking and hand tracking enabled. The Hololens 2 offers cutting-edge see-through holographic lenses with a field of view of 52°, a display resolution of 2048 x 1080 pixels per eye, and powerful tracking capabilities. The large field of view and the see-through lens of the Hololens 2 allow the users to preserve their natural spatial understanding. The higher resolution of the device allows a seamless blend of the virtual elements to the real world to create a superior immersion. Furthermore, Hololens eye tracking analyzes the unique feature of the eyes, such as interpupillary distance, and uses this data to generate personalized immersive experiences that prioritize the user's comfort and hologram alignment. To add natural and realistic walking animations, we decided to use motion capture to record walking animation. For this task, we used the Xsens full-body motion capture system and recorded multiple animations with different gait patterns at 60Hz.

\subsection {Avatar Selection}
\label{subsec:avatarselection}

The virtual elements are mainly comprised of a User Interface (UI) and an avatar. The user can select an avatar from a diverse group of avatars by swapping the UI up and down. This was designed to aid the user through complexity matching \cite{complexity-matching} and the proteus effect \cite{proteus:effect}. For example, there has been credible evidence that avatar appearance and body type influence the users \cite{vr:body1, vr:skincolor1}. To further extend the sense of realism to generate the illusory effect of body ownership \cite{vr:bodyRealism}, we used realistic humanoid avatars coupled with high-quality, motion-captured walk animations. 

Similarly, this is also designed to promote external focus and implicit learning, which are key motor learning principles. Traditional gait training systems heavily rely on explicit instructions from physical therapists. This promotes internal focus and diverts the patient's attention to the movements resulting in focused deliberate movements. However, natural movements happen as a result of automatic motor controls. Johnson et al argue that instructions that promote internal focus reduce automaticity and hinder learning and retention \cite{internalFocusReduce}. In addition, many credible and convincing studies support the theory that training gait with an external focus and implicit learning has better results such as the utilization of unconscious, fast, and reflexive control processes \cite{externalFocus1, externalFocus2, externalFocus3, externalFocus4, externalFocus5}. Therefore, we designed our system to provide external cues using avatars and promote external focus and implicit learning instead of providing explicit instructions. 

\subsection{UI Design}
\label{subsec:ui}

\begin{figure}[!t]
 \centering
 \includegraphics[width=1\columnwidth]{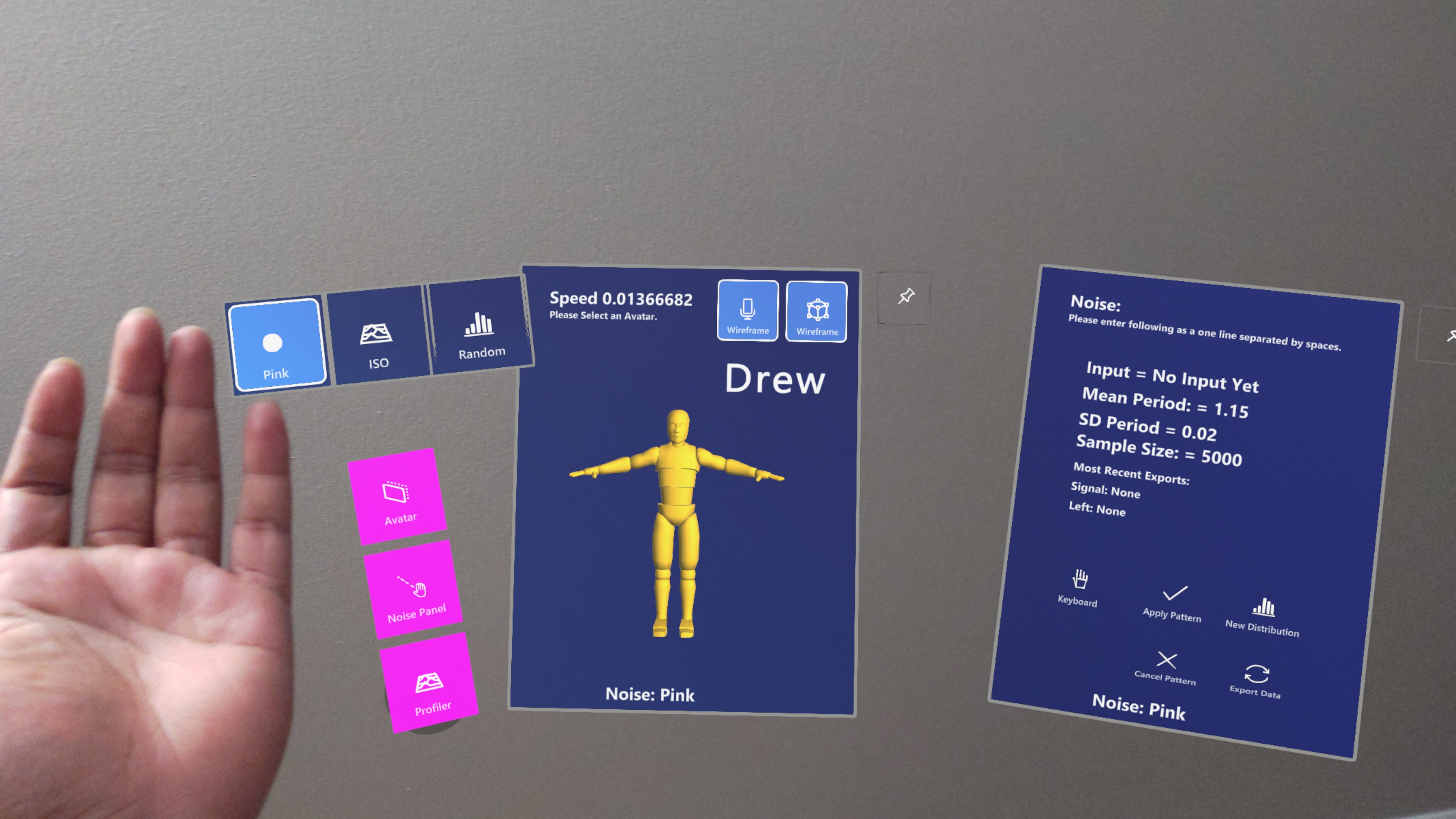}
 \caption{UIs of ARWalker. Upper left: Allows to select noise (described in detail in Section~\ref{subsec:noises}), Lower left: hand-bound menu to toggle the UI visibility, Middle: Allows to select an avatar, Right: Allows to generate custom noise for the patient.}
 \label{Figure:GaitAppUI}
\end{figure}

Holographic UI design differs from traditional UI design. The developers need to consider non-traditional variables such as environment light conditions, sound cues, and UI placements in dynamic environments. When designing our UIs we followed guidelines recommended by Microsoft for Hololens. Figure~\ref{Figure:GaitAppUI} shows the UIs that we have designed for ARWalker. 

\subsubsection{Impact of Colors}

Hololens 2 creates holograms by adding light to real-world light. As a result of additive displays, certain colors may appear differently on holographic displays. For example, white will appear bright, while black renders transparent. In other words, if one chooses a traditional simple background, such as a white background with black text, the whole background will be overwhelming and users will be able to see through the text. Similarly, in mixed reality, warm colors tend to stand out while cool colors recede into the background. Because of that, ARWalker, as several other mixed reality applications, uses dark colors for the background and white for the text elements. Microsoft guidelines for Hololens recommend the use of RGB values around (235, 235, 235) for white color and very dark grey RGB values for black color (16, 16, 16). We used these recommendations when designing our UIs and avatars. We used the MRTK default back plate with dark blue color when designing our UIs. 

\subsubsection{Materials}

Materials play a crucial role when rendering realistic holograms, since they provide proper visual characteristics that help the blending of virtual elements into physical space. Furthermore, materials provide visual feedback to users, so that they make sense of depth and distance, and can assist with dynamic lighting conditions. To handle these situations, we used the Mixed Reality Toolkit Standard shader (MRTK shader). MRTK shaders are designed, so that cross-platform development is facilitated, aesthetics are preserved, and performance is not sacrificed. 


\subsubsection{Other Factors}

We design our UIs with simplicity in mind. All the UIs are clearly labeled and have clear icons. Furthermore, our UIs have visual and auditory cues to provide feedback to the user. We designed our avatar UI to offer a similar experience to a modern phone. To select avatars, users have to swipe up and down. 
We also provide a quick hand-bound UI to toggle the visibility of the UIs. This allows the users to bring up necessary UIs quickly and then remove them from the view. 
We provide the option of pinning the UIs in a static position or having them follow the user dynamically. 

The UIs are designed not to block each others using the MRTK solver handles. Similarly, the UIs are placed in a position relevant to the head so that the user does not have to rotate their head in order to locate them. The UIs are guaranteed to stay within the user's field of view. Finally, we provide multiple input modalities to interact with UIs. It is possible for users to interact with them using hand rays, voice commands, and hand interactions from a distance, if needed. Each successful interaction provides visual and auditory cues confirming the interaction. 

\subsection {Design of Colored Noises}
\label{subsec:noises}




The gait of the avatar (stride times) follows different noise structures (pink, white, and isochronous). These noise structures have been found to represent properties of healthy and pathological human gait and we describe them below. 

\subsubsection{Pink Noise}

Over the past decades, researchers have become increasingly interested in studying the relationship between different events or values and their relationship over time in a series of data. Understanding the relationships and changes of the relationship between variables over a large timespan has benefits for a lot of areas of science, including human movement, medical sciences, and economics. Initially, researchers were concerned about short-term dependencies, which means that the current value depends upon a previous value or a few consecutive previous values. Over time, researchers have discovered that, in many cases, the value of the current value in a series is not simply determined by the value immediately preceding it, but also by the value from an earlier point in the series. The pattern may repeat at different time scales, such as hours, days, weeks, or even years. This means that the current value remembers the values that came before it, and this is called long-term memory, long-range dependence, fractal process, or 1/f noise. 

Researchers discovered that 1/f noise emerged in many complex systems and situations. For example, Kobayashi et al. discussed how 1/f noise has been found in the heartbeat series \cite{gait:heartbeat1f}. Similarly, Hausdorff et al. showed how 1/f noise emerges in the stride time series \cite{gait:stride1f}. Gilden et al. also provided credible evidence of how 1/f noise is associated with cognitive processes \cite{gait:cognitive1f}. Furthermore, Diniz et al. presented a solid analysis of 1/f noise and how it relates to motor control \cite{gait:theories1f}. The interest in 1/f noise peaked upon the discovery of its relationship with health. 
It is generally present in young, healthy systems performing a variety of easy tasks, but tends to disappear as the system ages or becomes ill \cite{gait:health1f}. In biomechanics and human movement, the gait of young and healthy individuals has been found to follow a pink noise distribution \cite{gait:pinkNoise}.

We used these principles to design pink noise, which is a signal with a power spectral density inversely proportional to the frequency. Based on the data collected during gait analysis, we generate a personalized pink noise pattern for each user. We then apply the noise to the avatar animation so that when the avatar walks, each gait cycle simulates a different speed according to the correct period of oscillation. The change in speed is directly related to the standard deviation of stride times obtained from an initial trial of self-paced walking for each participant. In other words, the speed does not exceed the range of speeds observed from the participant's self-selected cadence. Our pink noise module was developed with the intention of facilitating implicit learning through the use of the proteus effect and complexity matching techniques, so that participants are trained to follow the healthy gait exhibited by the avatar thus restoring their natural, healthy gait through the adoption of pink noise characteristics.

\subsubsection{White Noise}

As we age, our physiology deteriorates, and we tend to exhibit a white noise signature in things such as heartbeat and stride intervals. In terms of characteristics, white noise maintains equal power across all of its frequencies resulting in a static rhythm. A white noise signal generates a random pattern that is difficult to predict. ARWalker includes an option for the stride intervals of the avatar's gait follow a white noise distribution, which can be used to investigate how it affects young and healthy participants. 


The white noise function is based on a normally distributed random number. Our implementation utilizes the Box-Muller transformation to generate normally distributed random numbers. This transformation accepts two uniformly distributed deviates within the unit circle and transforms them to two independently distributed normal deviates based on the following equation:


\begin{equation} \label{Equ.3.1}	
y=\mu + \sigma * (\sqrt{ -2 \lg(X_1)} \cos( 2 \pi  X_2 ) )
\end{equation}
where $\mu = 0$, $\sigma = 1$, and $X_1, X_2$ are two uniformly distributed deviates within the unit circle.

\subsubsection{Isochronous Noise}

Isochronous noise 
is a constant value without variation. Participants with gait disorders can benefit from isochronous cues by establishing a stable and predictable rhythm that will enable them to synchronize their steps and maintain a steady gait pattern. Spaulding et al presented credible evidence that isochronous noise can be particularly effective in improving gait speed and stride length in individuals with Parkinson's disease, stroke, and other movement disorders \cite{gait:isoevidence}. Furthermore, isochronous noise can reduce the risk of falls, improve balance, and assist individuals in overcoming the freezing of their gait.
ARWalker offer the option for the stride times of the avatar's gait to follow an isochronous noise. At the beginning of the ARWalker operation, the user can enter a value and throughout the lifetime of the system, this value is maintained as a constant. 

\subsection{Coupling Noise with Animations}
\label{subsec:coupling}

In ARWalker, we adjust our animation in accordance with the frequency of the noise. In our motion-captured animations, a gait cycle (the interval between successive heel strikes of a single leg) will be completed within 1 second. However, we need to alter this animation so each gait cycle is completed with the appropriate duration from the selected noise. For example, if we have a sequence of durations that includes values $2.15, 1.28, 1.84, ..$ , the first gait cycle needs to be completed within 2.15 seconds, the second gait cycle within 1.28 seconds, the third gait cycle within 1.84 seconds, and so on so forth. This is a substantial challenge, since unity's traditional animations are predefined and hard to alter without breaking them. 

 To tackle this challenge, we separated a single gait cycle (with a length of 1.18 seconds) and used it as a baseline. We then manipulated the playback speed of the baseline gait cycle using the Unity animator speed component to create new gait cycles. Through this method, we were able to expand and shrink the length of the baseline gait cycle based on the values of the selected noise noise type to create a sequence of gait cycles that follow that noise type. 

\section{Evaluation}
\label{sec:evaluation}

In this section, we present our evaluation setup and results.


\subsection{Evaluation Setup}

We evaluated the performance of ARWalker using the Windows Performance Toolkit included in the Windows Assessment and Deployment Kit. As part of the Windows Performance Toolkit, there is a recording tool called Windows Performance Recorder (WPR), which is capable of recording  performance data about the Windows operating systems and applications (including applications running on a Hololens 2 headset). 
We created a custom profile to pull and record HoloLens 2 system performance data into an ETL trace file. To automate the analysis of data from multiple ETL files (each run of an experiment we conducted had a separate ETL file), we created a helper program using Java and wpaexporter. This program also allowed us to convert data from multiple ETL files into CSV files by iterating over them. Then using python, we created scripts that go through these data logs and visualize the Hololens 2 systems performance data.



For the purpose of our evaluation, we collected five traces of application behavior when using the pink noise and another five traces of application behavior when using the white noise. We did not collect traces for the ISO noise since it is a constant value. Our procedure for collecting the data is as follows. We started the application and then generated 5000 samples of noise using the mean value of 1.15 and standard deviation period value of 0.02. Our pilot study suggested that these values are more appropriate for human walking. After generating the noise, we applied them to the avatar animation system and walked with the avatar for five minutes while capturing the ETL traces. 
After gathering the traces, we export them into CSV files and we then quantify the CPU, GPU, memory, and disk consumption of the Hololens 2 headset while running the ARWalker application.


\subsection{Evaluation Results}

\subsubsection{CPU Performance}

The Hololens 2 CPU is mainly responsible for processing input, animations, physics, and other application logic. Figure~\ref{fig:CPU_AVG} shows the average CPU performance of our application when using both pink and white noise. Although the process of generating pink noise involves more mathematical operations than that of generating white noise, our results indicate that white noise consumes a slightly greater amount of CPU power than pink noise. The cause of this can be attributed to two main factors. When generating pink noise, it blocks other threads for a few seconds (3-4 seconds) for 5000 samples. This might be the reason behind the differences. We tried to address this freeze using coroutines in Unity. The basic idea behind the coroutine was to spread the calculations over several frames instead of one frame. We also tried to rewrite the function asynchronously. Both approaches complicated the code and introduced unpredictable elements, such as occasional crashes. Therefore, we decided not to use these and leave them as it is for the current version. Our plan is to use the Unity Job function to run the calculations in parallel in a separate thread after the initial patient testing is completed. A second reason might be the randomness of white noise. Compared to pink noise animation, white noise animation tends to move faster and slower in a random manner. Since the CPU handles the animation engine and animation state transitions, this might consume more CPU resources. 

\subsubsection{GPU Performance}
The Hololens 2 GPU handles various tasks such as pixel manipulation, handling draw calls, converting data into 2D and 3D shapes, and all the operations of the graphic pipeline. Our results as illustrated in figure~\ref{fig:GPU_AVG} show that pink noise consumes slightly more GPU power than white noise. However, it is worth noticing that, the consumption of GPU resources depends on many external factors. For example, we match the movement and rotation of the user to the movements and rotations of the avatar. This means if the user rotates too much or walks too fast, the GPU needs to redraw the UI elements and avatar to match the user's position and rotation. This can increase GPU resource consumption. Additionally, environmental factors also can affect the rendering of the virtual elements. For instance, in the case of an avatar being behind a table or covered by a ghost mesh, the GPU will not render the part of the avatar that is covered by the object. As a result, the GPU may receive fewer draw calls, which may result in fewer calculations being performed on the GPU. Therefore, the GPU calculations can vary between many external factors. 

\subsubsection{Memory Performance}
\begin{figure}[hbt!]
     \centering
     \begin{subfigure}[b]{0.49\textwidth}
         \centering
         \includegraphics[width=\textwidth, height=5cm]{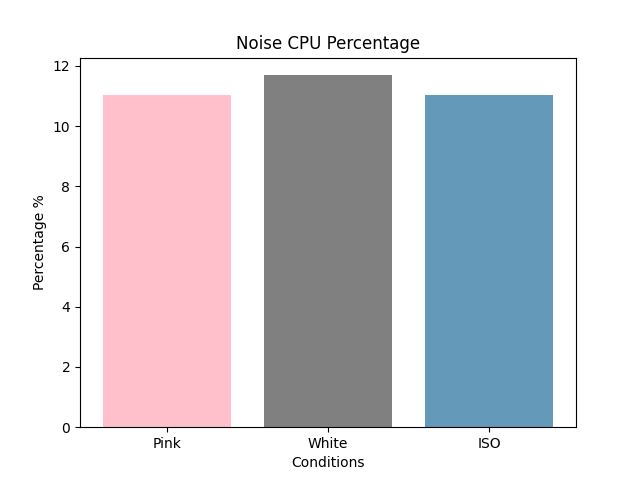}
         \caption{CPU Performance}
         \label{fig:CPU_AVG}
     \end{subfigure}
     \hfill
     \begin{subfigure}[b]{0.49\textwidth}
         \centering
         \includegraphics[width=\textwidth, height=5cm]{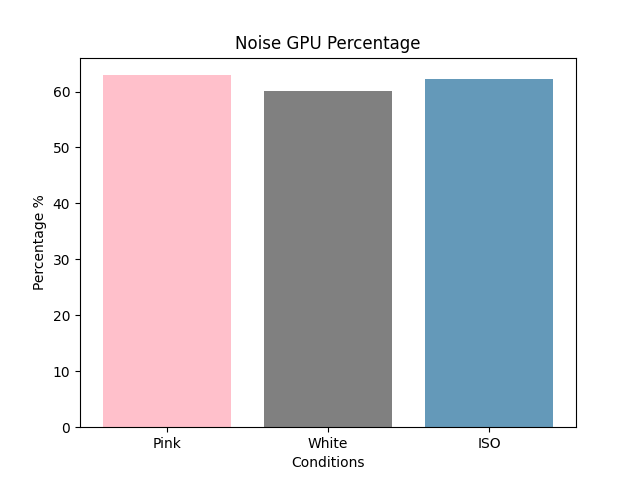}
         \caption{GPU Performance}
         \label{fig:GPU_AVG}
     \end{subfigure}
     \hfill
     \begin{subfigure}[b]{0.49\textwidth}
         \centering
         \includegraphics[width=\textwidth, height=5cm]{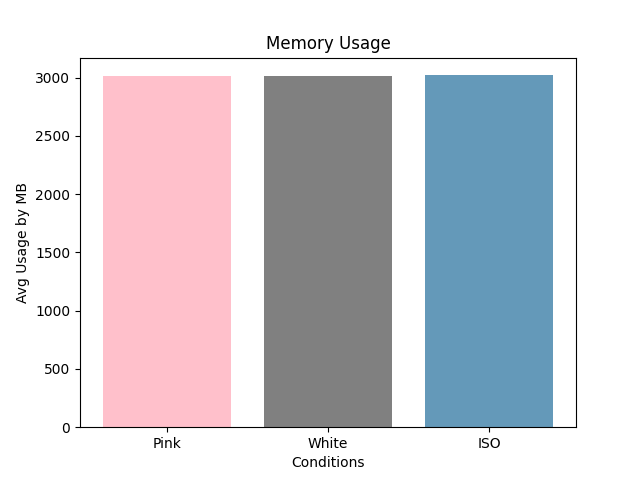}
         \caption{Memory Performance}
         \label{fig:Memory_AVG}
     \end{subfigure}
     \caption{System Performance Data}
     \label{fig:system_performance}
\end{figure}

Figure~\ref{fig:Memory_AVG} shows that both pink and white noise consume an equal amount of memory. This behavior is expected and there are two reasons behind this. When Hololens OS detects free available memory, it increases memory consumption to maximize efficiency. For example, in this kind of situation, Hololens cache more data in the memory in order to reduce the overhead of reading these data from storage. This means Hololens might use more memory than the application required to increase speed and efficiency.  

The second reason is when developing our application we used object pooling to reduce the cost of continuous memory allocation and deallocation. The concept of object pooling is a popular game development strategy where developers allocate more objects at the start of a game than necessary and then recycle these objects throughout the game's lifetime instead of creating and destroying objects.
Continuous creation and destruction of objects have higher performance overhead and object pooling can be used to avoid these high-cost operations.  As an example, in our application, if the user generates multiple noises with the same sample size, the list will be reused instead of allocating new memory every time. To conserve resources, many elements such as UIs and objects are reused throughout the program's lifetime. Furthermore, we used the Singleton design pattern for all the controllers. This prevents controllers from having more than one object throughout the lifespan of the program. This can be another factor of equal memory consumption. 

\subsubsection{Noise Generation Benchmark}
In addition to the generic application performance, we also evaluated the performance of the noise generation using the same methodology. After starting the traces and the application, the noise is generated, and the application and the traces are stopped. Table~\ref{table:noise_gen_benchmark} shows the benchmark data for noise generation. 
\begin{table}[ht]
	\centering
	\label{Band}
	\begin{tabular}{|l|l|l|l|}
		\hline
		NoiseType & CPU Core (avg) & GPU (avg) & Memory (avg) \\ \hline
		Pink & 10.95\% & 62.83\% & 3331.40 Mb \\ \hline
		White & 11.38\% & 72.86\% & 3332.88 Mb \\ \hline
  	ISO & 13.60\% & 64.77\% & 3099.20 Mb\\ \hline
	\end{tabular}
	\caption{Benchmark data of the noise generation.}
 \label{table:noise_gen_benchmark}
\end{table}

The noise generation results show slightly higher CPU core usage for the ISO noise compared to the pink and white noise.  Despite using the same environment, different environmental factors, including background noise, and light, may affect the application differently. Our application does not explicitly use cameras and microphones to generate noise, however, our voice command feature listens to background noise, and Hololens continuously analyzes the environment to ensure that the virtual element is blended with the real environment. CPU usage can be increased in different ways by these factors.    

GPU usage can also fluctuate differently depending on external factors. Hololens renders the virtual element relative to the user.  This means the user's position and rotation can affect the GPU performance. In our application, UIs are bound to stay in the user's field of view. Each time user changes the head position, it will place more draw calls to draw the UIs increasing the GPU resource consumption. 

The memory performance shows the expected behavior. Both pink and white noise uses mathematical libraries and generates intermediate calculations and store that in memory. Therefore, higher memory consumption is expected in both pink and white noise generation compared to the constant ISO value. 
\section{Discussion}
\label{sec:discussion} 

In this section, we discuss various considerations and challenges that we faced during the design and development of ARWalker. 

\subsection{Impact of Environmental Conditions and Other External Factors}

The Hololens 2 headset blends virtual elements into the real world with precision and realism. To do so, the headset needs to track the users in a space. Without proper tracking, the device fails to understand the environment and its relationship to the users. When this happens, it hurts the accuracy and stability of the avatars. For example, avatars may not appear properly or appear at odd locations of a user's field of view. This can negatively affect realism and disrupt the implicit learning that ARWalker relies on. The tracking performance of the device heavily relies on a combination of external factors and environmental conditions. We discuss external factors and conditions that may negatively impact ARWalker below. 

\subsubsection{Light} 

The most important environmental factor for ARWalker is light. ARWalker needs an even and sufficiently bright light to keep track of the user.
The depth perception of the headset also relies on the camera feed. We used this ability in ARWalker to ensure that the avatar is placed at a suitable distance from the user. 
Too dark or too bright light overwhelms the camera by making it unable to pick up information. Furthermore, an environment with uneven light (e.g., bright spots and dark spots, flickering lights) can confuse the camera by making it infer that a change of light is equal to a change of location. As avatars in ARWalker are rendered relative to the user's position, mistracking can result in avatars being rendered away from the user causing confusion and hindering the implicit learning process. 

Similarly, natural light can also cause instability in tracking. For example, when the lighting is greater than or equal to 10,000 lux, which is equivalent to outdoor lighting, Hololens 2 achieved between a 2\% and 3\% of contrast ratio. A low contrast ratio means that users are unable to accurately see the presence of virtual elements. This can produce different results. For instance, users may be able to see through the avatars or may not be able to see them at all. 

We have also found that uneven lighting, different light combinations (white and yellow light), and artificial and natural light combinations can negatively impact the operation of ARWalker. We tested ARWalker in a large indoor track field with a 1/8 mile distance. The track field has large windows for natural light and some parts of the field have yellow artificial light, while other parts have white artificial light. When we transitioned from an area with white light to an area with yellow light, we observed some unstable avatar behavior, such as flickering, and the avatar being rendered at the wrong locations. This case was more frequent when we approached areas that have both natural and artificial lights. To avoid this type of situation, Microsoft recommends maintaining a steady level of 500-1000 lux for optimal tracking results.


\subsubsection{Space}

To understand the space and track the position effectively, Hololens uses unique features in the space. These features can be anything as long as they have a unique shape or pattern. For example, a poster or symbol on a wall, a plant, or any other unique object can help the device to locate itself in the space. The device cannot track its position in a feature-poor space. In addition, if there are repeated features in the space, for instance, the same poster in multiple identical areas, can confuse the device. The Microsoft documentation suggests that creating unique (non-repetitive) patterns with lines of masking tapes can be a good solution to this issue. Similarly, two identical areas can also mislead the headset. In this case, the device cannot distinguish the different environments and infers that they are the same environment. Since ARWalker relies on tracking the user's position, rendering avatars at the appropriate locations can be impacted if the user is in feature-poor areas. 

On the other hand, a constantly and rapidly changing environment can also introduce tracking issues. When the number of moving objects increases, the quality of tracking decreases, as the headset has no stable features to locate against. This can cause ghost representation of meshes around the environment representing an object that is no longer there in reality. 



\subsection {Tradeoffs Among Different ARWalker Design Iterations}

In this section, we describe different versions of ARWalker that we initially developed (before developing the final version described in Section~\ref{sec:design}) and we discuss the challenges that we faced, which eventually led us to the development of the final version of ARWalker.

\subsubsection{Scene Understanding Version}

During our first iteration, we used the MRTK scene understanding component along with the Unity Navmesh component. Using scene understanding, we were able to extract floors, walls, and other objects in the environment in real time. Then, through the use of navmesh, we labeled walkable and unwalkable areas on the floor. The navmesh component also provides humanoid navigation functionality including path finding, so that avatars can walk in a natural manner avoiding obstacles in the environment. 

The version we created with MRTK scene understanding performed well in small and medium (up to 10x10 meter) environments. However, when running ARWalker in a large space (50x50 meters), the headset struggles to analyze the environment and crashes due to memory and processing constraints. 
A possible workaround for this that we plan to explore in the future is to offload camera frames to a server with a powerful GPU and send the processed data back to the headset~\cite{al2022reservoir, al2022promise, mastorakis2021networking}. The downside of this solution is it makes ARWalker dependent on network latency. Areas of poor WiFi coverage/connectivity may also impact the operation of ARWalker. 


\begin{figure}[!t]
\centering
\includegraphics[width=1\columnwidth]{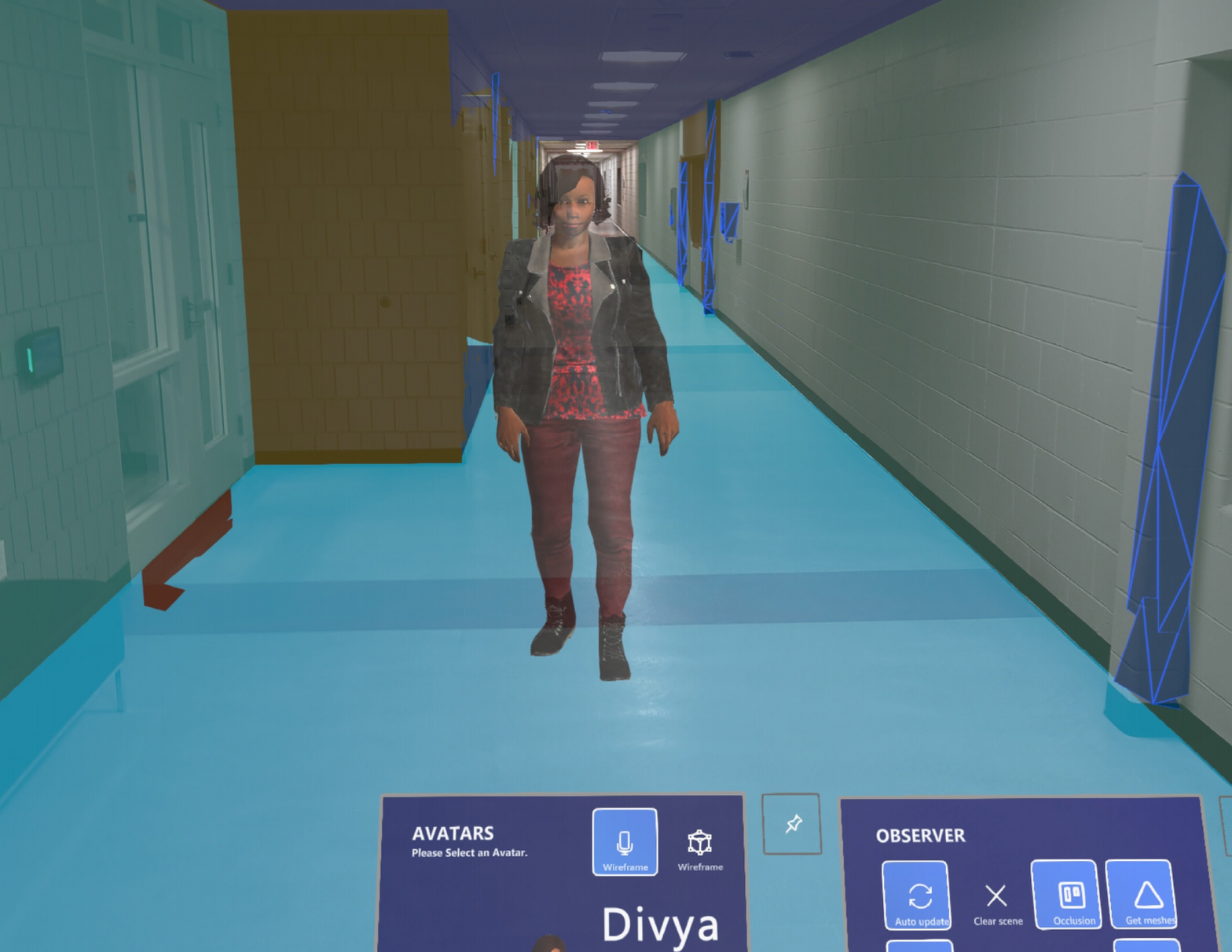}
\caption{An example of an avatar using scene understanding.}
\label{Figure:SUGhostMesh}
\end{figure}

\subsubsection{Waypoint Version}

During our second iteration, the design of ARWalker involved a mechanism that allowed users to define a path by placing waypoints. This approach can also be implemented through the Azure Spatial Anchor service. Spatial anchors allow the user to save points of interest, in this case, waypoints, on a remote server and to recall those points of interest from the headset. This method is easier to implement, but it also can introduce network latency issues and be impacted in cases of poor connectivity. 

This approach requires a calibration phase to define the desired path that the avatar will walk on by placing waypoints along this path. Furthermore, the headset defines the world position relative to the user's position. This is problematic because if waypoints are saved on a server and are loaded again on the headset, in order to guarantee that they are placed at the right locations of the avatar's path, users need to use the same starting position. Otherwise, the waypoints might be placed at incorrect locations of the avatar's path. 

While testing this version, we noted that waypoints often drifted around the intended positions. Generally, this occurs for three reasons. When the headset starts collecting data about the world for the first time, it estimates the distance between different physical and virtual elements. The headset adjusts these estimations to actual distance as more data is analyzed. For example, when we place a waypoint in the context of ARWalker, the headset might estimate that the distance between the waypoint and the nearest wall is 0.3 meters. As it learns more about the environment, it may update its initial estimate to an actual value, such as 0.4 meters. As a result, the headset moves the waypoint 0.1 meters away from the wall to be in line with the actual distance. The second reason is incorrect tracking because of external factors such as light. The third reason is when virtual elements are placed outside a 5-meter diameter from the headset, the headset struggles to track and maintain the positions of these elements. Since our track field is 1/8 miles long, the headset struggles to keep track of waypoints placed along the track field as users walk around the track field. Due to these issues and during our third iteration, we eventually designed ARWalker as we described in Section~\ref{sec:design}.

\begin{figure}[!t]
\centering
\includegraphics[width=1\columnwidth]{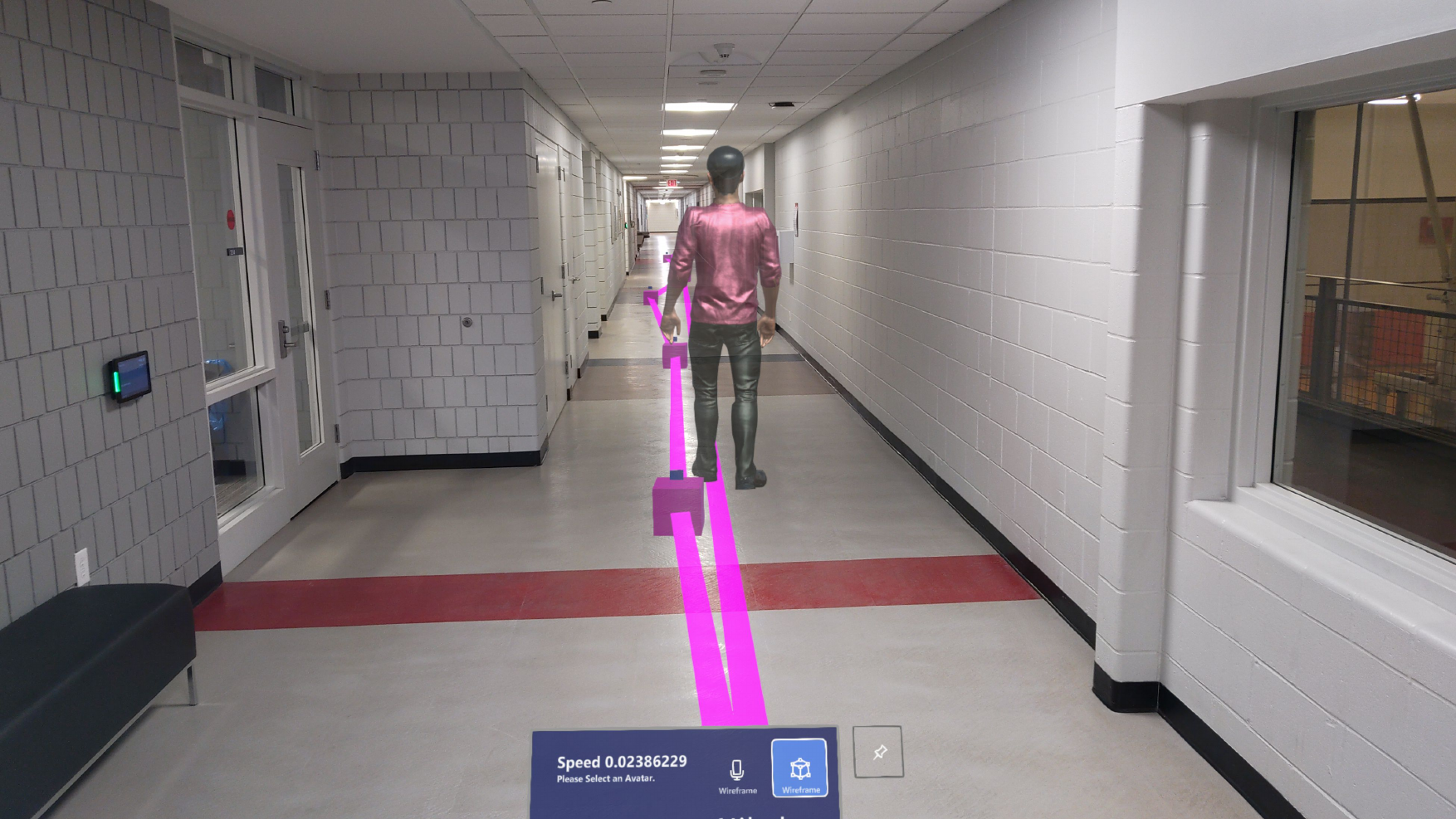}
\caption{An example of an avatar walking between waypoints.}
\label{Figure:SUGhostMesh}
\end{figure}

\section{Conclusion and Future Work}
\label{sec:conclusion}

In this paper, we presented ARWalker, an AR application that aims to offer gait training and rehabilitation activities for older adults and populations with diseases. We discussed various design and development considerations and challenges we faced. Our evaluation of the systems performance of ARWalker demonstrated that it can run without consuming significant resources on a Hololens 2 headset. In the future, we plan to: (i) focus on the rehabilitation and clinical aspects of ARWalker and use it to conduct gait rehabilitation studies on older adults and other populations; and (ii) explore mechanisms to offload parts of the data processing happening on the headset (e.g., scene understanding) to remote servers.

\section*{Ethical Considerations}

The work presented in this paper has been approved by the Institutional Review Board (IRB) of the University of Nebraska (IRB\#:0009-22-EP).

\section*{Acknowledgments}

This work is partially supported by the National Science Foundation (awards CNS-2104700, CNS-2306685, CNS-2016714, and CBET-2124918), ACM SIGMOBILE, and the National Institutes of Health (award NIGMS/P20GM109090).

\bibliographystyle{unsrt}
\bibliography{sections/references}

\section*{Biographies}

\vskip -5.0\baselineskip plus -4fil

\begin{IEEEbiographynophoto}{Pubudu Wijesooriya}
(pwijesooriya@unomaha.edu) is an M.S. student in Computer Science at the University of Nebraska at Omaha. 
His research interests include computer systems and extended reality technologies.
\end{IEEEbiographynophoto}





\vskip -5.0\baselineskip plus -4fil

\begin{IEEEbiographynophoto}{Aaron Likens}
(alikens@unomaha.edu) received his Ph.D. from the Arizona State University. He is an Assistant Professor in Biomechanics at the University of Nebraska Omaha. His interests include nonlinear dynamics. 
\end{IEEEbiographynophoto}



\vskip -5.0\baselineskip plus -4fil

\begin{IEEEbiographynophoto}{Nick Stergiou} (nstergiou@unomaha.edu) is the Distinguished Community Research Chair in Biomechanics and the Director of the Center for Research in Human Movement Variability, University of Nebraska Omaha. 
His research focuses on understanding variability in human movement. 
\end{IEEEbiographynophoto}

\vskip -5.0\baselineskip plus -4fil

\begin{IEEEbiographynophoto}{Spyridon Mastorakis}
(mastorakis@nd.edu) is an Assistant Professor in Computer Science and Engineering at the University of Notre Dame. He received his Ph.D. in Computer Science from UCLA in 2019. 
His research interests include network systems, edge computing, IoT, and security.
\end{IEEEbiographynophoto}

\vskip -5.0\baselineskip plus -4fil

\end{document}